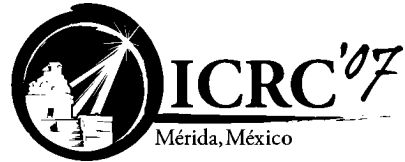

# The ANTARES Detector: Electronics and Readout

M. CIRCELLA[1] ON BEHALF OF THE ANTARES COLLABORATION
[1]*Istituto Nazionale di Fisica Nucleare (INFN) – Sezione di Bari, via Amendola 173, I-70126 Bari, Italy*
*Marco.Circella@ba.infn.it*

**Abstract:** ANTARES is a large volume, deep-sea neutrino telescope currently under construction off the coast of Toulon, France. The apparatus consists of an array of 900 large-area photomultipliers arranged on 12 detection lines. 5 of these lines are in data acquisition since January 2007, and the schedule is to complete the apparatus construction in early 2008. In this paper, we illustrate the main features of the electronics and data acquisition systems of the apparatus.

## Introduction

ANTARES is an underwater neutrino telescope under construction at a depth of 2475 m in the Mediterranean Sea. The installation site is at about 40 km off the coast of Toulon, France. The control station is installed in Institut Michel Pacha in La Seyne Sur Mer, close to Toulon.

The apparatus consists of an array of 900 photomultiplier tubes (PMTs) by which the faint light pulses emitted by fast charged particles propagating in the water may be detected. Based on such measurements, ANTARES will be capable of identifying neutrinos of atmospheric as well as of astrophysical origin. In addition, the detector will serve as a monitoring station for geophysics and sea science investigations.

Since January 2007, 5 of the 12 detection lines of the apparatus are in data acquisition. The plan is to complete the construction of the apparatus in early 2008.

For an introduction to the scientific aims of the ANTARES experiment, the reader is referred to the dedicated presentation at this Conference [1] and to the online documentation [2].

## The ANTARES Apparatus

The detector consists of an array of 900 large-area photomultipliers (PMTs), Hamamatsu R7081-20, enclosed in pressure-resistant glass spheres to constitute the optical modules (OMs) [3], and arranged on 12 detection lines. An additional line will be equipped with environmental devices.

Each line is anchored to the seabed and kept in a vertical position by a top buoy. The minimum distance between two lines is of about 60 m.

Each detection line is composed by 25 storeys, each equipped with 3 photomultipliers oriented at 45° with respect to the vertical. The storeys are spaced by 14.5 m, the lowest one being located about 100 m above the seabed. The storeys are connected by an electro-mechanical cable, which is equipped with 21 optical fibers for digital communications.

From the functional point of view, each line is divided into 5 sectors, each of which consists typically of 5 storeys. Each storey is controlled by a Local Control Module (LCM), and each sector is provided with a modified LCM, the Master Local Control Module (MLCM), which maintains the data communications between its sector and the shore. A String Control Module (SCM), located at the basis of each line, interfaces the line to the rest of the apparatus.

Each of these modules consists of an aluminum frame, which holds all electronics boards connected through a backplane and is enclosed in a water-tight titanium cylinder. The internal frame has been designed so as to transfer the heat generated inside the module toward the external container, so as to protect the electronics from over-heating risks.

The contents of electronics boards in each container are chosen depending on the specific configurations of the modules. The minimum contents are: a power box, a data acquisition board, 3 front-end boards for the 3 optical modules, a clock board, and a compass board, which provides positioning measurements from a system of compass and inclinometers. Additional boards may be used for control of the further calibration or monitoring devices connected to the module or for the specific MLCM/SCM functions.

A network of submarine cables supports the power distribution and all communications for detector control and data acquisition. All offshore communications take place over optical fibers.

A 50 km electro-optical cable links the telescope to the onshore station. This cable is equipped with copper conductors for power distribution (AC with sea-return) and 48 optical fibers for digital communications. The cable is terminated by an underwater junction box, which can support connections for up to 16 lines.

## Electronics and Data Acquisition (DAQ)

### Requirements and constraints

The most compelling requirements that the electronics and data acquisition systems for a complex installation such as ANTARES have to fulfil are a large bandwidth for data transmission, low power consumption and a very high reliability.

The data bandwidth requirements are mainly set by the level of optical background due to the radioactive decay of $^{40}K$ and to the bioluminescence activity, which in the installation site amounts to a minimum rate of 70 kHz per PMT.

The typical data flow expected from the full detector is of the order of 7.5 GB/s. ANTARES is designed to sustain such a high data rate by means of the Dense Wavelength Division Multiplexing (DWDM) technique described in the following.

Power consumption should be kept at minimum for two main reasons: power is provided by the onshore station, therefore increasing the power needs will affect the structure (and consequent cost) of the main electro-optical cable as well as of the entire offshore network. Secondly, the power dissipated offshore will affect the temperature balance inside the electronics containers, and higher temperatures may negatively affect the lifetimes of the components.

Finally, reliability and long-term functionality isues of the detector have received a large attention in the design and implementation of the ANTARES telescope. Access to the instrumentation is impossible, once the detector is installed, unless an entire line is recovered. This procedure, however, is time-consuming and expensive.

Several solutions have been adopted in order to improve the overall reliability of the detector. Firstly, the network of underwater communications has been designed so as to minimize the impact of single point failures: the most sophisticated functions are performed by components distributed along the whole detector, while mainly passive operations are performed at the 'network nodes' such as the junction box and the SCMs. Secondly, only very well known and tested components are used in the offshore electronics. Thirdly, a detailed test and qualification scenario has been developed for the line production, and each step of further integration toward the construction of each line is marked by a qualification test of the instrumentation being assembled. In addition, all of the equipment to be used in the offshore electronics underwent an Accelerated Stress Test, in order to get rid of the infant mortality and to increase the MTTF of the surviving components.

Finally, the Collaboration has set up a detailed Quality Control plan, which allows to trace the life of individual components and ensures that all elements in the ANTARES lines are conform to their design specifications.

### The front-end electronics

The full-custom Analogue Ring Sampler (ARS) has been developed to perform the complex front-end operations. This chip samples the PMT signal continuously at a tunable frequency up to 1 GHz and holds the analogue information on 128 switched capacitors if a threshold level is crossed. The information is then digitized, in response to a trigger signal, by means of an integrated dual 8-bit ADC. Optionally the dynamic range may be increased by sampling the signal from the last dynode.

A 20 MHz reference clock is used for time stamping the signals. A Time to Voltage Converter (TVC) device is used for high-resolution time measurments between clock pulses.

The ARS is also capable of discriminating between simple pulses due to conversion of single photoelectrons (SPE) from more complex waveforms. The criteria used to discriminate between the two classes are based on the amplitude of the signal, the time above threshold and the occurrence of multiple peaks within a time gate. Only the charge and time information is recorded for SPE events, while a full waveform analysis is performed for all other events.

The ARS chips are arranged on a motherboard to serve the optical modules. Two ARS chips, in a 'token ring' configuration, perform the charge and time information of a single PMT. A third chip provided on each board is used for triggering purposes.

The settings of each individual chip can be remotely configured from the shore.

## Trigger

Two levels of trigger, completely configurable from the shore, have been implemented (L0 and L1). The lowest level of trigger, L0, is generated by any ARS chip when the input signal exceeds the threshold. As it is preferred that all data filtering occurs onshore, this is the routine trigger condition for the apparatus.

However, in peculiar conditions, for instance in periods of high optical background, the amount of data to transfer to shore may be limited by requiring simple coincidence conditions among the signals from the optical modules of the same storey, L1. In this case, the PMT data are momentarily stored by the ARS into a pipeline from which they can be retrieved and sent to the DAQ board if a request arrives from it within a preset time window.

## DAQ and slow control

A single processor is in charge of the data acquisition and slow control (SC) in each storey: the choice has been for the low-power RISC Motorola MPC860P running the VxWorks Operating System. The MPC860P features a 100 Mb/s ethernet controller, 4 Serial Communication Controllers (SCC) and a Serial Peripheral Interface (SPI). The ethernet port is used for external data/SC communications, while two of the SCC ports are used for local (i.e., internal to the module) slow control communications on two RS485 serial bus, with MODBUS protocol.

The interface between the ARSs and the DAQ processor is provided by an FPGA (Altera APEX20K200E), which receives the data from the front-end chips, temporarily stores them into internal buffers and then moves them to the DAQ memory (a SDRAM).

The processor is in charge of retrieving the data from the DAQ memory and sending them, through an ethernet switch located in the MLCM, toward the shore station. The FPGA code is stored on a dedicated EEPROM.

## Long-distance ethernet communications

The long-distance communications are organized with a star topology. Independent communications with any line take place along different optical fibers of the main cable. The data from the different sectors of a line are sent toward shore implementing a DWDM technique with standard ITU optical wavelengths with 400 GHz frequency spacing. The selected wavelengths lie in the region between 1535 nm and 1560 nm. In each sector the gateway to the network is performed by the MLCM, which features a DWDM laser for data transmission to the shore and a 1 Gb/s ethernet switch for communications with the DAQ processors inside the MLCM itself and the other LCMs of the sector.

The data coming from the different sectors are merged by a DWDM multiplexer in the SCM and then injected into a single fiber toward the shore station. On the shore, the reverse process is implemented with a passive DWDM demultiplexer.

Similarly, the network supports the data transfer from the shore to the offshore electronics.

## Onshore data processing

Depending on the operational mode of the detector, each line is capable of producing data at a rate between 50 Mb/s to 1 Gb/s. A fast data processing system is used onshore for selecting the data to be recorded. This system consists of an exapandable farm of PCs. The number of PCs needed for data acquisition from the full apparatus will not exceed 100. The data flow through this system is managed through the IP addressing of the ethernet protocol: the offshore modules choose the destination address where to send their data packets depending on the time stamp. Consequently, each onshore node processes the data collected in the whole detector in defined time windows (10 ms of duration, typically). In order to avoid network congestion, the deliveries of data packets from different modules to the same node are staggered.

Several trigger algorithms are executed on the onshore PC farm. In addition, a continuous record of up to 100 s of data are temporarily kept in memory in such a way that they can be saved to disk in case of an external trigger signal, such as a GRB alarm.

More details on the data communication and processing systems are reported elsewhere [4].

## Calibration devices

The reconstruction of the particle events is based on the interpretation of the charge and time measurements of the light pulses detected by the different optical modules. Hence, the pointing accuracy of the telescope depends critically on how well the OM positions are known and the signal detection times are measured.

ANTARES is equipped with a hybrid positioning system. Each offshore electronics container is equipped with a tiltmeter and a compass, giving information on the 3-axis orientation of the storey. In addition, an acoustic triangulation system provides measurements of the propagation times, hence 3-D distances, from acoustic transceivers located at the bases of the lines to hydrophones properly distributed along the strings. The shape of each line is then reconstructed by performing a global fit of all this information.

Optical (LED and laser) beacons are distributed at proper location in the apparatus and flashed at known times, so as to provide reference signals for time alignment of the measurements by the different optical modules.

For more details on the calibration systems of the apparatus, see [5].

## Clock

The clock system has been designed to provide a low-jitter common clock reference for the whole detector. A 20 MHz master clock signal is generated onboard and distributed to all electronics containers offshore by means of dedicated optical fibers of the network of underwater cables described previously. Digital data may be superimposed on the reference clock, thus providing a means of distributing common synchronised orders to the whole apparatus or to an addressable part of it.

The clock network supports bidirectional transmissions. The return path is used for measuring the propagation delay between the onshore clock system and the offshore electronics, for time calibration purposes, as well as for sending status and monitoring information to shore.

The onshore clock is synchronised to the GPS time with an accuracy of 100 ns.

## Conclusions and Outlook

With more than 350 optical modules from 5 lines in acquisition since January 2007, the ANTARES apparatus is already the largest neutrino telescope in operation in the northern hemisphere. All subsystems behave nominally and data taking is proceeding smoothly.

The next connection operation is expected in Fall 2007, when 3 new lines will be connected. The apparatus will reach its final configuration by early 2008.